\documentclass{elsart}
\usepackage{graphicx}
\usepackage{amssymb}
\usepackage{amsmath}
\usepackage{bm}

\begin{document}
\begin{frontmatter}

\title{Final State Interaction Effects in the $e^+e^-\to N\bar{N}$ Process near Threshold}

\author{V.F. Dmitriev and A.I. Milstein}
\address{Budker Institute of Nuclear Physics, 630090 Novosibirsk, Russia}

\begin{abstract}
We use the Paris nucleon-antinucleon optical potential  for
explanation of experimental data in the process  $e^+e^- \rightarrow
p\bar p$ near  threshold. It is shown that the cross section and the
electromagnetic form factors are very sensitive to the parameters of
the potential. It turns out that final-state interaction due to
slightly modified absorptive part of the potential allows us to
reproduce available experimental data. We also demonstrated that the cross section in $n\bar{n}$ channel is larger than that in $p\bar{p}$ one, and their ratio is almost energy independent up to 2.2 GeV.
\end{abstract}


\end{frontmatter}

\section{Introduction}
At present,  QCD can not describe  quantitatively the low-energy
nucleon-antinucleon interaction, and various  phenomenological
approaches have been suggested in order to explain numerous
experimental data , see , e.g., Refs.
\cite{pot82,bonn91,bonn95,paris82,paris94,Partial94} and recent
reviews \cite{KBMR02,KBR05}. However, parameters of the models still
can not be extracted with a good accuracy  from the experimental
data \cite{Rich95}.

Very recently, renewed  interest in low-energy nucleon-antinucleon
physics has been stimulated by the experimental observation of a
strong enhancement of decay probability at low invariant mass of
$p\bar p$ in the processes  $J/\Psi\to \gamma p\bar p$ \cite{BES03},
$B^+\to K^+p\bar p$ and $B^0\to D^0 p\bar p$
\cite{Belle02a,Belle02b,BaBar05}, $B^+\to \pi^+ p\bar p$ and $B^+\to
K^0 p\bar p$ \cite{Belle04c}, $\Upsilon\to \gamma p\bar p$
\cite{CLEO06}. One of the most natural explanation of this
enhancement is final state interaction of the proton and antiproton
\cite{Ker04,Bugg04,Zou04,Loi05,Sib05,Sib06}.

A similar phenomenon was observed  in the investigation of the
proton (antiproton) electric, $G_E(Q^2)$, and  magnetic, $G_M(Q^2)$,
form factors  in the process $e^+e^- \to p\bar p$
\cite{Bardin94,Armstrong93,Aubert06}. Namely, it was found that the
ratio $|G_E(Q^2)/G_M(Q^2)|$ strongly depends on $Q^2=4E^2$ (in the
center-of-mass frame) in the narrow  region of  the energy $E$  near
the  threshold of $p\bar p$ production. Such  strong dependence at
small $(E-M)/M$ is related to the interaction of proton and
antiproton at large distances $r>>1/M$. Therefore, it is possible to take one of the nucleon-antinucleon interactions determined in 
\cite{pot82,bonn91,bonn95,paris82,paris94,Partial94} for describing the final state interaction
 in the process  $e^+e^- \rightarrow
p\bar p$ in order to explain the experimental data. In the present paper, we use the Paris nucleon-antinucleon
optical potential $V_{N\bar N}$  which has the form
\cite{paris82,paris94}:
\begin{equation}\label{VNN}
V_{N\bar N} = U_{N\bar N}-i\, W_{N\bar N} \, ,
\end{equation}
where the real part $U_{N\bar N}$  is the  $G$-parity transform of
the well established Paris $NN$ potential for the long- and
medium-ranged distances ($r\gtrsim 1\mbox{fm}$), and some
phenomenological part for the short distances. The absorptive part
$W_{N\bar N}$ of the optical potential takes into account the
inelastic channels of $N\bar N$ interaction, i.e. annihilation into
mesons. It is essential at short distances and depends on the
kinetic energy of the particles. Our knowledge of $W_{N\bar N}$ is
essentially more restricted than that of $U_{N\bar N}$. Therefore,
we hope that experimental data for the cross section of the process
$e^+e^- \rightarrow p\bar p$ can significantly diminish the
uncertainty in $W_{N\bar N}$.

\section{Amplitude of the process}
Using the standard definition of the Dirac form factors $F_1(Q^2)$ and $F_2(Q^2)$ 
\cite{LL} of the proton we write, in the nonrelativistic approximation, the amplitude of $N\bar{N}$
pair production  near  threshold as follows (in
units $e^2/Q^2$):
\begin{eqnarray}\label{2}
T_{\lambda\mu} &=& \bm{\epsilon}_\lambda^* \Biggl\{
\frac{1}{3}\bigl[(2E+M)F_1(Q^2)+\frac{E^2+2ME}{M}F_2(Q^2)\bigr]
{\bf e}_\mu  \nonumber\\
&&+\frac{EF_2(Q^2)-MF_1(Q^2)}{3M(E+M)}\bigl[3({\bf p}'\cdot{\bf
e}_\mu){\bf p}'-{\bf p}'^2{\bf e}_\mu\bigr]\Biggr\}\, ,
\end{eqnarray}
where ${\bf e}_\mu$ is a virtual photon polarization vector, and
$\bm{\epsilon}_\lambda$ is the spin-1 function of $N\bar{N}$ pair.
Two tensor structures in Eq.(\ref{2}) correspond to the s-wave and
d-wave production amplitudes. The total angular momentum of the
$N\bar{N}$ pair is fixed by a production mechanism. The functions
$F_1(Q^2)$ and $F_2(Q^2)$ are related in a standard way with the electric and the
magnetic form factors of the proton $G_E=F_1+\frac{Q^2}{4M^2}F_2$
and $G_M=F_1+F_2$.  Near threshold, it is more convenient to work
with the Dirac form factors, because at threshold $G_E=G_M$. The
amplitude (\ref{2}) already includes  effects of final state
interaction. Therefore, the form factors in Eq.(\ref{2}) should have
a pronounced $Q^2$ behavior near threshold. Our aim is to single out
these effects. In order to do that, let us introduce the amplitude
$\tilde{T}_{\lambda\mu}$ which has the form of Eq.(\ref{2}) but with
the replacement $F_1\to\tilde{F}_1$ and  $F_2\to\tilde{F}_2$ , where
the form factors $\tilde{F}_{1,2}$ do not account for the effect of
final state interaction. So, the difference between $T_{\lambda\mu}$
and $\tilde{T}_{\lambda\mu}$  is in their  $Q^2$ dependence due to
the form factors. In  $\tilde{T}_{\lambda\mu}$,  the only scale is
the nucleon mass $M$ (which is still too small to use perturbative
QCD). Therefore, near threshold, where the final state interaction
is important, the form factors $\tilde{F}_{1,2}$ are  smooth
functions of $Q^2$ and can be treated  as  phenomenological
constants.

Now we can construct the  amplitude of  $N\bar{N}$ pair production
in a certain isospin channel $I=0,1$ using the  wave function
$\mathbf{\Phi}^{I(-) \dagger}_{{\bf p}\lambda}({\bf p}')$ of the
$N\bar{N}$ pair in  momentum space:
\begin{eqnarray}\label{4}
T_{\lambda\mu}^I & =& \int \frac{d^3p'}{(2\pi)^3}\mathbf{\Phi}^{I(-)
\dagger}_{{\bf p}\lambda}({\bf p}')
\Biggl\{\frac{1}{3}\bigl[(2E+M)\tilde{F}^I_1+\frac{E^2+2ME}{M}\tilde{F}_2^I\bigr]
{\bf e}_\mu \nonumber\\
&&+\frac{E\tilde{F}^I_2-M\tilde{F}^I_1}{3M(E+M)}\bigl[3({\bf
p}'\cdot {\bf e}_\mu){\bf p}' -{\bf p}'^2{\bf e}_\mu\bigr]\Biggr\}\,
.
\end{eqnarray}
The Fourier transform of the function $\mathbf{\Phi}^{I(-)
\dagger}_{{\bf p}\lambda}({\bf p}')$, which is the wave function
$\mathbf{\Psi}^{I(-)\dagger}_{{\bf p}\lambda}({\bf r})$ of the
$N\bar{N}$ pair in  coordinate space,  is the solution of the
Schr\"odinger equation
\begin{equation}
\mathbf{\Psi}^{I(-)\dagger}_{{\bf p}\lambda}({\bf r})\hat{H}=(E-M)
\mathbf{\Psi}^{I(-)\dagger}_{{\bf p}\lambda}({\bf r})\,,\quad
\hat{H}=\frac{{\bf p}^2}{M}+V_{N\bar{N}}\,,
\end{equation}
where $V_{N\bar{N}}$ is given by Eq.(\ref{VNN}). Note that
$\mathbf{\Psi}^{I(-)\dagger}_{{\bf p}\lambda}({\bf r})$ is the left
eigenfunction of the bi-orthogonal set of eigenfunctions of the
non-Hermitian operator $\hat{H}$.
 Its asymptotic behavior at large distances is
\begin{equation} \label{6}
\mathbf{\Psi}^{I(-)\dagger}_{{\bf p}\lambda}({\bf r})\approx
e^{-\imath {\bf p}\cdot {\bf r}}+\tilde{f}^I\frac{e^{\imath
pr}}{r}\,\,.
\end{equation}
For $p\bar{p}$ production, we have
$T_{\lambda\mu}^p=T_{\lambda\mu}^1+T_{\lambda\mu}^0$, while for
$n\bar{n}\;\; T_{\lambda\mu}^n=T_{\lambda\mu}^1-T_{\lambda\mu}^0$.

\subsection{Coupled channels basis}
In the presence of tensor forces, the states with angular momentum
$L$ and $L+2$ are coupled, while the total angular momentum $J$ is
conserved. As a result, the final-state wave function
$\mathbf{\Psi}^{I(-) \dagger}_{{\bf p}\lambda}({\bf r})$ can be
represented in the form
\begin{eqnarray}\label{9}
&&\mathbf{\Psi}^{I(-)\dagger}_{{\bf p}\lambda}({\bf
r})=\sum_{JM\alpha} D^{\alpha IJM*}_\lambda \mathbf{\Psi}^{\alpha
IJM\dagger}({\bf r})+  \sum_{JM}F^{IJM*}_\lambda
v^I_J(r)\mathbf{Y}^J_{JM}({\bf n})\, ,
\end{eqnarray}
where $\bm n=\bm r/r$ ,
\begin{equation} \label{7}
\mathbf{\Psi}^{\alpha IJM}({\bf r})= u^{\alpha
I}_J(r)\mathbf{Y}^{J-1}_{JM}({\bf n}) + w^{\alpha
I}_J(r)\mathbf{Y}^{J+1}_{JM}({\bf n})\quad(\alpha=1\,,2)\, ,
\end{equation}
 and
\begin{equation} \label{8}
v^I_J(r)\mathbf{Y}^J_{JM}({\bf n})
\end{equation}
are three  independent solutions of the Schr\"{o}dinger equation
having  total angular momentum $J$. In Eqs.(\ref{7}) and (\ref{8}) ,
$\mathbf{ Y}^L_{JM}(\theta,\phi)$ is the vector spherical harmonic,
which is an eigenfunction of the operators $\mathbf{L}^2,\mathbf{J}^2$, and $J_z$,
where $\mathbf{J}=\mathbf{L}+\mathbf{S}$ and $S=1$. In order to find
the coefficients $D^{\alpha IJM}_\lambda$ and $F^{IJM}_\lambda$ in
Eq. (\ref{9}), we substitute  the asymptotics of the functions $
u^{\alpha I}_J(r)$ , $w^{\alpha I}_J(r)$ and $v^I_J(r)$ at large $r$
in Eq. (\ref{9}) and use Eq.(\ref{6}). Then we obtain
\begin{eqnarray}\label{10}
D^{\alpha IJM}_\lambda = 4\pi \sum_{Lm}\imath^L\langle
JM|Lm1\lambda\rangle Y^*_{Lm}(\hat{\bf p})A^{I\alpha}_L\,
,\nonumber\\
 F^{IJM}_\lambda = 4\pi\sum_m\imath^J\langle
JM|Jm1\lambda\rangle Y^*_{Jm}(\hat{\bf p})A^{I}_J\,,
\end{eqnarray}
where the coefficients $A^{I\alpha}_L$ and $A^{I}_J$ are related to
incoming flux in the asymptotics  of the radial functions
$u^{I\alpha}_J(r), w^{I\alpha}_J(r)$ and $v^I_J(r)$. Making the
Fourier transform of  the expansion Eq.(\ref{9}), substituting the
result in Eq.(\ref{4}),  and performing the integration over the
angles of the vector $\bm p'$, we obtain
\begin{eqnarray}\label{11}
T_{\lambda\mu}^I &=&\int
\frac{p'^2dp'}{(2\pi)^3}\Biggl\{\sqrt{4\pi}\Bigl[\frac{1}{3}(2E+M)\tilde{F}^I_1
+ \frac{E^2+2ME}{3M}\tilde{F}_2^I\Bigr]
\nonumber\\
&&\times\sum_\alpha D^{\alpha I1\mu *}_\lambda f^{I\alpha}_1(p')
+\sqrt{2\pi}\frac{E\tilde{F}^I_2-M\tilde{F}^I_1}{3M(E+M)}p'^2\sum_\alpha\,
D^{I\alpha 1\mu *}_\lambda g^{I\alpha}_1(p')\Biggr\}\, ,
\end{eqnarray}
where $f^{I\alpha}_J(p')$ and $g^{I\alpha}_J(p')$ are the
corresponding radial parts of the wave functions in  momentum space.
Integrals over $p'$ are related to the values of radial wave
functions at $r=0$.
\begin{eqnarray} \label{12}
u^{I\alpha}_1 (0)=\frac{1}{2\pi^2}\int p^2\,f^{I\alpha}_1(p)dp\quad
,\quad w^{I\alpha''}_1(0) =-\frac{1}{15\pi^2}\int
p^4\,g^{I\alpha}_1(p) dp\, ,
\end{eqnarray}
where $w^{I\alpha''}_1(0)$ is the second derivative at $r=0$. The amplitude Eq.(\ref{11}) becomes
\begin{eqnarray}  \label{13}
T_{\lambda\mu}^I
&=&\frac{1}{\sqrt{4\pi}}\Big[\frac{1}{3}(2E+M)\tilde{F}^I_1
+\frac{E^2+2ME}{3M}\tilde{F}_2^I\Big]\sum_\alpha D^{\alpha
I1\mu *}_\lambda u^{I\alpha}_1(0)\nonumber\\
&&+\frac{5}{4}\frac{E\tilde{F}^I_2-M\tilde{F}^I_1}{M(E+M)\sqrt{2\pi}}\sum_\alpha
D^{I\alpha 1\mu *}_\lambda w^{I\alpha''}_1(0)\,.
\end{eqnarray}
Using Eq.(\ref{10}) for the coefficients $D$, we can present the
amplitude in the form
\begin{equation}   \label{14}
T_{\lambda\mu}^I = G_0^I \delta_{\lambda\mu} -\sqrt{4\pi}G_2^I
\langle 1\mu|2m\,1\lambda\rangle Y^*_{2m}(\hat{\bf p}),
\end{equation}
where
\begin{eqnarray}\label{15}
G_0^I&=&\left[\frac{1}{3}(2E+M)\tilde{F}^I_1+\frac{E^2+2ME}{3M}\tilde{F}_2^I\right]
\sum_\alpha A^{I\alpha}_0 u^{I\alpha}_1(0)\nonumber\\
&&+\frac{5\sqrt{2}}{4}\frac{E\tilde{F}^I_2-M\tilde{F}^I_1}{M(E+M)}
\sum_\alpha A^{I\alpha}_0 w^{I\alpha''}_1(0)\,,\\
G_2^I&=&\left[\frac{1}{3}(2E+M)\tilde{F}^I_1+\frac{E^2+2ME}{3M}\tilde{F}_2^I\right]
 \sum_\alpha A^{I\alpha}_2
u^{I\alpha}_1(0)\nonumber\\
&&+\frac{5}{2\sqrt{2}}\frac{E\tilde{F}^I_2-M\tilde{F}^I_1}{M(E+M)}
\sum_\alpha A^{I\alpha}_2 w^{I\alpha''}_1(0)\,.
\end{eqnarray}
The observable electric and magnetic form factors are expressed in terms of $G^I_0$ and $G^I_2$ in the following way
\begin{equation} \label{17}
G_E^I=G^I_0+\sqrt{2}\,G^I_2\frac{Q}{2M},\;\;
G_M^I=G^I_0+\frac{1}{\sqrt{2}}G^I_2\,.
\end{equation}
\section{$p\bar{p}$ production}

The proton form factors are the sum of isoscalar and isovector form
factors. The differential cross section for $p\bar{p}$ production is
\begin{eqnarray}\label{unpolar}
\frac{d\sigma}{d\Omega}= \frac{\alpha^2\beta
C}{4Q^2}\left[|G^p_M(Q^2)|^2(1+\cos^2{\theta})+\frac{4M^2}{Q^2}|G^p_E(Q^2)|^2\sin^2{\theta}\right]\,,
\end{eqnarray}
where $\beta=v/c$ and $C$ is the Coulomb distortion factor. We
omitted here effects of Coulomb-nuclear interference. Using the form
factors $\tilde{F}^I_i$ as free parameters, we fit the observed
energy behavior of the cross section up to 100 MeV of proton kinetic
energy. We found that using the original Paris $N\bar{N}$ potential \cite{paris99}
it is impossible  to reproduce the energy behavior of the cross
section. The cross section falls down
 too steeply  becoming  more than one order of magnitude smaller than the observed one
  at 10 MeV of proton c.m. kinetic energy. We verified that possible smooth dependence of the form factors $\tilde{F}^I_i$ on Q
 in a narrow energy region near threshold (for instance, $\tilde{F}^I_i\propto 1+b(Q^2-4M^2)/8M^2$ with $b\sim 1$), 
 which is not related to final state interaction, does not change this result. 
 The result is also insensitive to variation  of most parameters of the potential. Variation  
 of the energy independent parameters is absorbed by the fitting parameters $\tilde{F}^I_i$. 
  However, if we modify
  the only parameter, the energy dependence of absorptive part of the
  triplet potential $W_{N\bar{N}}$, decreasing it by a factor of $8 \div 10$, we obtain a good fit
  of the cross section (see Fig.1).  Modification of the energy
 dependence of the real part of the potential is not so important. Note
 that the values  of the parameters, which are responsible for the
 energy dependence of the potential, are not well known. In the two versions of
  the Paris $N\bar{N}$ potential, \cite{paris94} and \cite{paris99}, these parameters
 differ from each other  up to factor 2. In order to clarify the importance of the energy dependence of the absorptive potential, we calculated S-matrix elements and corresponding phase shifts for $J=1$, $S=1$ at different energies. It turns out that the phase shifts obtained with our strong modification of the absorptive potential differ only slightly as compared to those obtained in \cite{paris94}. The only noticeable modification appeared in the energy dependence of the parameter $\eta$ directly related to absorption. In contrast, $|\mathbf{\Psi}^{\alpha I1M}(0)|^2$, see Eq.(\ref{7}), is much more sensitive to this modification.  Therefore, the uncertainty in determination of the parameters of energy dependence in the absorptive part of the triplet potential from the scattering data is apparently larger than factor two, and it is necessary to use also another data. 
In the process $e^+e^-\rightarrow p\bar{p}$ we have a unique situation where the quantum numbers of $p\bar{p}$
pair are fixed, $J=S=1$. However, the absorptive part of $N\bar{N}$ effective potential is not universal and may be different in different scattering channels, as it takes place in Nijmengen  potential \cite{nijm94}. Thus, the modification of the energy dependence in the absorptive part of the potential for $J=S=1$ channel does not allow us to make any conclusions on other channels and a new fit of absorptive potential in these channels should be performed.

Using the parameters obtained from the fit of the
cross section, we simultaneously reproduce  the energy behavior of
the form factors ratio $|G_E/G_M|$ (see Fig. 2).  Emphasize that the form factor $G_E$ differs from the form factor $G_M$ due to the contribution of D-wave only, see Eq.(\ref{17}). Therefore, the strong energy dependence of the ratio $|G_E/G_M|$ clearly indicates the importance of D-wave even in the vicinity of threshold.  Having obtained the amplitudes for the two isospins, we have calculated the cross section and the ratio $|G_E/G_M|$ for the process $e^+e^-\rightarrow n\bar{n}$. The corresponding results are shown in Figs 1,2 by the dashed lines. It is seen that the final state interaction leads to strong enhancement of both quantities in $n\bar{n}$ channel as well. It is interesting that the cross section in $n\bar{n}$ is larger than that in $p\bar{p}$ channel, and their ratio is almost energy independent up to 2.2 GeV.

Very recently, the final state interaction in  $e^+e^-\rightarrow p\bar{p}$ has been discussed in Ref. \cite{sibnew}. The authors of the paper have presented many arguments in favor of importance of final state interaction in $p\bar{p}$ production near threshold. Using J\"ulich model for $N\bar{N}$ interaction they have calculated the contribution of $^3S_1$ partial wave to the $e^+e^-\rightarrow p\bar{p}$ cross section near threshold. However, since they have neglected the $^3D_1$ partial wave contribution, they have not been able to reproduce the ratio $|G_E/G_M|$ which is equal to unity in their approximation.

\begin{figure}[h]
\includegraphics[width=0.5\textwidth ,height=7cm]{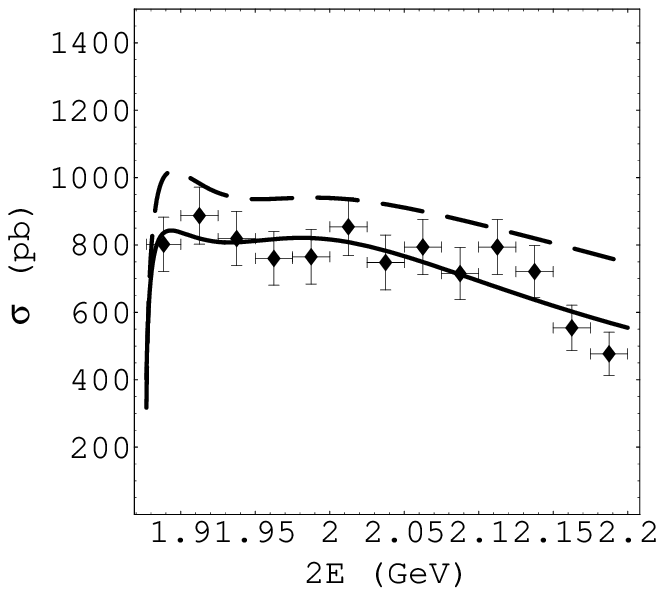}
\hfill
\includegraphics[width=0.5\textwidth ,height=7cm]{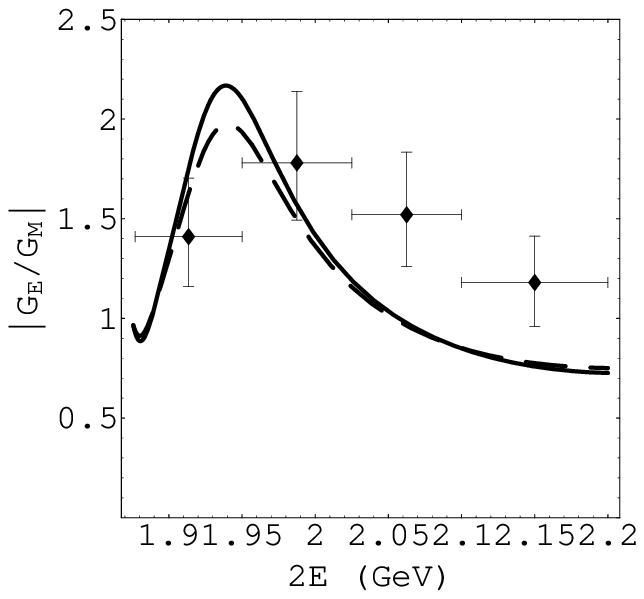}
\\
\parbox[t]{0.46\textwidth}{\caption{Fit of the cross section of $p\bar{p}$ production (solid line). Data are from Ref.\cite{Aubert06}. Dashed line:  the cross section for $n\bar{n}$ production}}
\hfill
\parbox[t]{0.46\textwidth}{\caption{Ratio of $|G_E^p/G^p_M|$ (solid line). Data are from Ref.\cite{Aubert06}. Dashed line:  the ratio for a neutron.}}
\end{figure}

In summary, we calculated the effects of final state interaction in
the reactions  $e^+e^-\rightarrow p\bar{p}$ and $e^+e^-\rightarrow n\bar{n}$ near threshold. We found
that the last version of the Paris $N\bar{N}$ potential \cite{paris99} does not reproduce the
energy dependence of the observed cross section. A smooth dependence of the form factors $\tilde{F}^I_i$ on Q
 in a narrow energy region near threshold, 
 which is not related to final state interaction, does not change this result. Variation of the
parameters  responsible for the energy dependence of the  imaginary part of the potential in the channel with $J=S=1$, and $L=0,2$
allows us to reproduce both the energy dependence of the cross
section an the ratio $|G_E/G_M|$ for $p\bar{p}$ production.
We obtained that the cross section in $n\bar{n}$ channel is larger than that in $p\bar{p}$ one, and their ratio is almost energy independent up to 2.2 GeV.
\section*{Acknowledgements}
 The authors appreciate discussions with V.P.~Druzhinin, G.V.~Fedotovich, A.A.~Sibirtsev, E.P.~Solodov, and V.M.~Strakhovenko.
The work was supported  in part by RFBR Grant No. 05-02-16079.

\end{document}